\documentclass[11pt]{article}

\usepackage{amssymb,amsthm,amsfonts,psfig,epsfig,axodraw}
\def\gappeq{\mathrel{\rlap {\raise.5ex\hbox{$>$}} {\lower.5ex\hbox{$\sim$}}}}
\def\lappeq{\mathrel{\rlap{\raise.5ex\hbox{$<$}} {\lower.5ex\hbox{$\sim$}}}}

\def\beq{\begin{equation}} \def\eeq{\end{equation}} \def\bea{\begin{eqnarray}}
\def\eea{\end{eqnarray}}\def\bq{\begin{quote}} \def\eq{\end{quote}}
\def\nn{\nonumber}

\begin{document} 

\pagestyle{empty} 
\begin{flushright}  SACLAY-T02/130\end{flushright}  

\vskip 1.5 cm    
\def\thefootnote{\fnsymbol{footnote}}
\begin{center}  
{\large \bf LEPTON FLAVOUR VIOLATION \protect\footnote{Proceedings
of SUSY02, 10th International Conference on {\it Supersymmetry and Unification of Fundamental
Interactions}, 17-23/06/'02, DESY, Hamburg, Germany.}}     
\end{center}  
\vspace*{5mm} 
\centerline{\bf Isabella Masina}
\vskip 0.5 cm
\centerline{\em Service de Physique Th\'eorique 
 , CEA-Saclay}
\centerline{\em F-91191 Gif-sur-Yvette, France} 
\vskip 1.5 cm   

\centerline{\bf Abstract}  

After a summary of the motivations for searching flavour violating decays 
and electric dipole moments in the leptonic sector, I briefly sketch the present status of 
the limits on lepton-slepton misalignement 
and discuss the potentialities of the planned experimental improvements. 
Candidates for theories beyond the Standard Model have to satisfy all these experimental constraints; 
I will discuss in particular how the supersymmetric SM extended with the see-saw faces the
flavour problem.  

\vspace*{1cm} 
\pagestyle{plain}


\def\thefootnote{\arabic{footnote}}
\setcounter{footnote}{0}

\section{Motivations}

There are many motivations for searching lepton flavour violating (LFV) decays - 
like $\tau \rightarrow \mu \gamma$, $\mu \rightarrow e \gamma$ -, 
fermion electric dipole moments (EDM) and additional contributions to fermion magnetic dipole moments: \\
\noindent {\bf 1.} lepton family number ($L_i$)  violation is already a well established fact; \\
\noindent {\bf 2.} such observables, if measured - and there will be significant improvements in the experimental
sensitivity in future - represent clean smoking guns for new physics; \\
\noindent {\bf 3.} they already are a severe test for theories beyond the Standard Model (SM), 
in particular for low energy supersymmetry. 

Actually, to explain why such processes have not already been observed is one of the major problems
of low energy supersymmetry: the {\it flavour problem}. 
In the following, I shall firstly briefly review the present constraints on the leptonic soft supersymmetry breaking 
terms and discuss the impact of the planned experimental improvements.
It is convenient to encode such constraints on ${\cal L}_{soft}$ in the form of upper bounds
on mass-insertion $\delta$'s.

In this way it is possible to test many extensions of the SM supplemented with low energy supersymmetry.  
The constraints from both the quark and leptonic sector of ${\cal L}_{soft}$ are so severe that the easiest
way out is to consider the possibility of family blind soft terms. For the SM, then, all the above processes 
and their analog in the quark sector would be extremely suppressed.
Nevertheless, even with universal soft terms, some extensions of the SM could give observable effects.
This is the case for supersymmetric grandunified theories in the quark sector, as addressed 
by  A. Masiero in his talk. In the leptonic sector, however, it is not even necessary to invoke a stage of 
grandunification to dispose of efficient beyond the SM interactions: 
it is quite likely that the desert below $M_{GUT}$ suggested by gauge coupling unification 
is actually populated by three generations of right-handed neutrinos and their interactions:
the see-saw mechanism.  
To estimate the prediction for LFV decays and EDM in the case that the SM is extended with the see-saw
is an intersting task: it represents the minimal amount of contribution that could be present. 
If, in addition, there is a stage of grandunification, since there is no reason for a 
conspiracy between contributions of different nature, such processes should 
eventually be enhanced. In the following, I will develop this issue in some detail.

\subsection{Basic Motivation n.1: $L_i$ Violation }

That $L_i$ is violated at low energy we are pretty sure: it is even more violated than $B_i$.
Indeed, neutrino oscillation experiments have shown that it is maximally violated because 
in the flavour basis where charged leptons are diagonal \footnote{Here and in the following, diagonal matrices 
will be denoted with a hat.} the MNS matrix which diagonalises light neutrinos, thus expressing how much 
$L_i$ is violated in the {\it effective} theory, 
\beq
m_\nu^{eff}=U_{MNS}^*  {\hat m}_\nu U_{MNS}^\dagger
\eeq
has turn out to possess two large angles, $\theta_{23}, \theta_{12} \sim \pi / 4$. Curiously enough, now the puzzle is 
to explain why the third angle is smaller than $\theta_C \approx .22$, which measures the amount of $B_i$
violation at low energy.

The most elegant explanation of the smallness of neutrino masses is the see-saw mechanism \cite{seesaw}:
\bea
{\cal L}_{h.e.} & \ni & \frac{1}{2} {\nu^c}^T {\hat M} {\nu^c} + {\nu^c}^T 
\underbrace{V_R {\hat y}_\nu V_L}_{= y_\nu} \nu H_u^0 \\
                    {\scriptstyle l.e.}   &    \searrow  &   \nn\\
                       &                    & y_\nu^T \frac{1}{\hat M} y_\nu v_u^2= m_\nu^{eff}  
\label{mnueff}
\eea 
In the {\it fundamental} theory at high energy, the amount of $L_i$ violation corresponds to the largeness
of the mixing angles of $V_{R}, V_{L}$, the unitary matrices which diagonalise $y_\nu$ in the basis where 
$M$ and charged leptons are diagonal.
Notice that $V_L$ (and not $U_{MNS}$) is the true analog of $V_{CKM}$ in the leptonic sector. 
Clearly, $L_i$ violation at low energy, i.e. $U_{MNS} \neq \mathbb{I}$, implies that $V_R$ and $V_L$ cannot both
be $\mathbb{I}$, but we cannot guess the magnitude of their angles. 
So, the question "is $V_L$ small like $V_{CKM}$ or large like $U_{MNS}$?" is not academic.

It is not possible to find an answer by just looking at neutrino phenomenology, namely at $m_\nu^{eff}$.
As can be realised by simple parameter counting, one cannot extract $V_R$, $V_L$.
It is better to separate our ignorance from our knowledge by recognizing 
that the extraction of the combination  
\beq
\frac{1}{\sqrt{ \hat M}} V_R {\hat y} V_L v_u 
\label{comb}
\eeq 
from 
\beq
m_\nu^{eff}=( U_{MNS}^*  { \sqrt{\hat m}_\nu}  R ) ( R^T{ \sqrt{\hat m}_\nu}   U_{MNS}^\dagger ) = 
(\frac{1}{\sqrt{\hat M}} V_R {\hat y} V_L v_u )^T  (\frac{1}{\sqrt{ \hat M}} V_R {\hat y} V_L v_u )
\eeq
is ambiguous up to $R$, an orthogonal complex matrix \cite{c&i} . 
Thus, $R$ parametrizes our ignorance in the sense that it encodes six physical informations that are 
present at high energy and that are subsequently lost when integrating out right-handed neutrinos \cite{lms2}.
To further disentangle the matrices in eq. (\ref{comb}), one should for example know $\hat M$.
As also appears from the parameter counting, the two sets below are absolutely equivalent in describing the 
see-saw mechanism. However, as will be clarified in the following, the physical meaning of a certain 
quantity can be better interpreted using one set than the other. 
\bea
~~~~~\{~~\underbrace{{\hat m},~~U_{MNS}}_{\rm known},~~\underbrace{R,~~{\hat M}}_{\rm unknown}~~\} 
\div  \{~~\underbrace{{\hat y},~~V_{L},~~V_R,~~{\hat M}}_{\rm fundamental}~~\}~~~~ \label{sets} \\
        {\rm real}     ~~~~~~~~~~~~~    3 ~~~~~3~~~~~~~~3~~~~~3~~~~~~~~~~~~~3 ~~~~~3~~~~~3~~~~~3~~~~~~~~~\nn\\
       {\rm phases}  ~~~~~~~~~~~~~~~~~~3~~~~~~~~3~~~~~~~~~~~~~~~~~~~~~~~~~~1~~~~~5~~~~~~~~~~~~~~~~\nn
\eea  

It is interesting to wonder what is the physical meaning of the informations contained
in $R$, which acts on $\nu$ and $\nu^c$ mass eigenstates at left and right respectively.
It turns out that it is a dominance 
\footnote{The idea of dominance has been firstly suggested and developed from 
the point of view of model building \cite{dom}.} 
matrix  \cite{lms2}: 
its pattern reflects the way in which each heavy right-handed neutrino contributes in determining
each light neutrino mass scale.
Consider for instance the heaviest light neutrino mass scale, $m_3$.
Due to the orthogonality of $R$,
\beq
m_3 = \sum_i R_{3i}^2 m_3 ~~~~,
\eeq
so that $m_3$ is determined by a sum of three contributions ${\cal C}_{3i} \equiv R_{3i}^2 m_3$.
It turns out that, once $U_{MNS}$ is fixed, 
each ${\cal C}_{3i}={\cal C}_{3i}(y_{ki}^2/M_i)$, i.e. is a function of $M_i$ and its couplings 
to the light neutrinos $y_{ki}$ only. $R_{3i}^2$ thus represent the weights of the $\nu^c$ in 
determining $m_3$. Analogous considerations hold for the various $m_j$. 
So, if a certain $\nu^c$ gives the dominant contribution in determining a certain eigenvalue $m_j$, 
correspondingly this information is encoded in the structure of the jth row of $R$. For instance, 
if $\nu^c_k$ mostly determines the scale of $m_j$, we will have $R_{jk}>>R_{jh}$, where $h \neq k$. 
The interpretation in terms of weights is straightforward for the rotational part of $R$.  
One has however to be careful because of the possible presence of the three boosts 
(controlled by the three phases). In the latter case the weights $R_{3i}^2$ are not necessarily 
positive nor smaller than 1. So, the situation is more subtle but, even including the phases of $R$, 
the interpretation in terms of dominance still holds.
The presence of a structure in $R$ is important. 
For hierarchical neutrinos for instance, if the
third row of $R$, which determines $m_{atm}$, is structureless, one cannot account at same time 
for the large atmospheric mixing and the experimental value of $r \equiv (\Delta m^2_{sol}/\Delta m^2_{atm})^{1/2}$ 
without the introduction of a tuning between the Yukawas at the level of $r$. 
With a proper structure in the third row of $R$, the latter requirements are instead naturally fulfilled.    

To proceede one needs other observables. 
Recently, a lot of attention has been given to the possibility of having baryogenesis through 
leptogenesis \cite{leptog}. 
Here I will just make some comments. It is interesting to notice that, in the case of hierarchical right-handed
neutrinos, say $M_1<M_2<M_3$, defining $R_{ij}= |R_{ij}| e^{i \phi_{ij}/2}$, the $\epsilon$ asymmetry
can be written by using the first set of parameters in eq. (\ref{sets}) and it is simply given by \cite{msetal} 
\beq
\epsilon = \frac{3}{16 \pi} \frac{-|R_{11}|^2 \Delta m_{sol}^2 \sin\phi_{11}+ |R_{31}|^2 \Delta m_{atm}^2 \sin\phi_{31}}
{|R_{11}|^2 m_1 +|R_{21}|^2 m_2 +|R_{31}|^2 m_3 } \frac{M_1}{v^2} ~~~~.
\label{eps}
\eeq  
With this parametrization, we can see mathematically what one would have expected by intuition, namely
that the physical quantities involved in determining $\epsilon$ are just $M_1$, the spectrum of light-neutrinos, 
$\hat m$, and the first column of $R$, which expresses the composition of $\nu^c_1$ in terms of the light neutrino 
masses $\hat m_i$. Since there is not at all trace of $U_{MNS}$,
from this formula one can immediately realise that $\epsilon$ (=leptogenesis) and $m_\nu^{eff}$ (=neutrino
phenomenology) depend on two independent sets of $CP$ violating phases \footnote{See for instance \cite{diffapp} 
and references therein for different approaches to this subject.}, 
respectively those of $R$ and those of $U_{MNS}$.
Notice also that if $V_L \rightarrow U_{MNS}$, i.e. we postulate that low energy left mixings arise
mostly from those at high energy, then $R \rightarrow \mathbb{I}$ and $\epsilon$ vanishes.
Leptogenesis can however provide just one number. To proceede further one need other
informations. 

\subsection{Basic Motivation n.2: New Physiscs "Smoking Guns" }

Let's have a look at the SM as an effective theory: the operators of dimension $d>4$ represent new physics
and could be computed if we knew the fundamental theory beyond the SM. The only one which has been observed
is the $d=5$ operator responsible for neutrino masses, eq. (\ref{mnueff}); 
however, the $d=6$ operator
\beq
\frac{1}{\Lambda^2} {\bar \psi} \sigma^{\mu \nu} (\gamma_5) \psi F_{\mu \nu} \phi 
\eeq
potentially induces LFV, EDM and additional contributions to MDM.  
Some upper limits on these observables and their probable future improvements
are displayed in table 1. In the SM they are predicted to be much smaller than
the present and future bounds. Observation of some of them would thus represent
a very clean smoking gun for the presence of new physics beyond the SM.

\begin{table}[!ht]
\begin{center}
\begin{tabular}{|c||c|c|} 
\hline & & \\
& present \cite{exppres} & planned \cite{expfut} \\ & &  \\ \hline  \hline &  & \\ 
$d_e$ & $ < 1.5~ 10^{-27} $ e cm & $< 10^{-29(-32)}$ e cm  \\ &  &  \\  \hline  &  & \\ 
$d_\mu$ &  $< 10^{-18} $ e cm   &  $ < 10^{-24(-26)}$ e cm  \\ &  &  \\  \hline  &  & \\ 
BR($\mu \rightarrow e \gamma$) & $ < 1.2~ 10^{-11} $ &    $< 10^{-14}$  \\ &  &  \\  \hline  &  & \\ 
BR($\tau \rightarrow \mu \gamma$) &  $<1.1~ 10^{-6} $  &  $ < 10^{-9}$(?)   \\  && \\  \hline   
\end{tabular} \\ \vspace{.2 cm}
Table 1
\end{center}
\end{table}

\subsection{Basic Motivation n.3: Test for Theories beyond the SM}

The upper limits provide the third motivation, namely to test theories beyond the SM.
Let me focus on the see-saw:
in the non supersymmetric case, these operators are suppressed by powers of 
the scale of $L$ violation and the effects are completely out of the future sensitivity;
with low energy supersymmetry, on the contrary, these operators are suppressed by powers of $m_{susy}$, 
due to graphs with loops of virtual supersymmetric particles. 
The effects should be so large that the supersymmetric flavour problem arises:
the only explanation of why such processes have not yet been found is to assume very specific
patterns of soft supersymmetry breaking terms at high energy, 
namely that fermions and sfermions have to be nearly diagonal in the same basis.

\section{Constraints on ${\cal M}^2_{\tilde \ell}$}

In the basis where charged fermions are diagonalised, non observation of LFV, EDM and additional
contributions to MDM in the leptonic sector, constrains the $6 \times 6$
hermitian matrix of soft sfermion masses, whose mixings represent the amount of fermion-sfermion
misalignement.
For practical purposes it is convenient to work in the mass insertion approximation
(see for instance \cite{mass_ins} and references therein): one assumes nearly degenerate sfermions so that the experimental
upper bound on a certain observable can be directly translated into upper bounds on some of the $\delta$'s 
defined in this way
\beq
{\cal M}^2_{\tilde \ell} = \left( \matrix{ m_{LL}^2 & m_{LR}^2 \cr m_{RL}^2 & m_{RR}^2  }\right) = 
{\bar m}_{\tilde f}^2 \left( \matrix{ \mathbb{I} + \delta_{LL}^2 & \delta_{LR}^2 \cr \delta_{RL}^2 & \mathbb{I}+\delta_{RR}^2  }\right) ~~~.
\eeq
The constraints on the $\delta$'s are so many that, if one is patient enough to tackle them in a systematic way,
one ends up with a sleptonarium \cite{ms_arium}, a catalogue of upper bounds on $\delta$'s.
These bounds can be for instance represented in the $(M_1, m_{\tilde e_R})$ 
plane for fixed $\tan \beta$ values assuming for definiteness gaugino universality 
and $\mu$ from radiative electroweak symmetry breaking (deviations from these assumptions are easily estimated). 
Besides the encyclopedical motivation, such a collection is interesting because
the combined analysis of all the limits allows to extract additional informations. 
In the following I will briefly sketch this issue (see \cite{ms_arium} for more details).

Non observation of the decay $\ell_i \rightarrow \ell_j \gamma$ puts upper bounds on the absolute value
of four $\delta_{ij}$:  $|\delta^{LL}_{ij}|$, $|\delta^{RR}_{ij}|$, $|\delta^{LR}_{ij}|$, $|\delta^{RL}_{ij}|$.
There are two possible contributions to EDM: flavour conserving (FC) and flavour violating (FV).
If no conspiracy between them is at work, the experimental bound can be put on each: 
$d_i^{exp} > d_i^{FC},~ d_i^{FV}$.  
The FC contribution bounds ${\cal I}m(\delta_{ii}^{LR})$. The FV contribution, which
could be even bigger than the FC one if there is a fermion $j$ heavier 
than the $i$, bounds the following four combinations of $\delta$'s:
\beq
{\cal I}m( \delta_{ij}^{LL}\delta_{ji}^{LR}) ~~~~,~~~~{\cal I}m( \delta_{ij}^{LR}\delta_{ji}^{RR}) ~~~~,~~~~
{\cal I}m( \delta_{ij}^{LL}\delta_{jj}^{LR}\delta_{ji}^{RR} )~~~, ~~~{\cal I}m( \delta_{ij}^{LR}\delta_{jj}^{RL}\delta_{ji}^{LR} ) ~~.   
\label{4im}
\eeq
Consider $d_\mu$ and let's have a look at the informations that can be obtained by combining 
the various bounds. 
Since the $|\delta_{23}|$ are independently constrained from $\tau \rightarrow \mu \gamma$, $d_\mu^{FV}$ 
cannot exceed a certain value.
For instance, a bound at the level of $10^{-8}$ on the $BR(\tau \rightarrow \mu \gamma)$ 
would imply that the FV contribution to $d_\mu$ is not bigger then 
$\sim 10^{-23}$ e cm, well below the present experimental bound on $d_\mu$ but in the range of 
the planned sensitivity. 
So, if $d_\mu$ where observed above $\sim 10^{-23}$ e cm, we could conclude that we are measuring the 
FC contribution; if it where observed below, its origin - if FV or FC - would remain unknown.
Similar remarks can be done for $d_e$.

\section{$\delta$'s as Tests for Theories Beyond the SM}

The interest of collecting the bounds on the $\delta$'s is that they can be easily compared with
the prediction of theories. 
In general we can distinguish two contributions to the $\delta$'s: one already present at the scale $\Lambda$
where boundary conditions are imposed and one which can be potentially developed when running 
from $\Lambda$ to $m_{susy}$:
\beq
\delta_{ij}=\delta_{ij}^{(0)}+\delta_{ij}^{(rad)}~~~~.
\eeq
Since there is no reason for contributions of such different origin to conspire, the above limits on the
$\delta$'s can be applied to both.

\begin{figure}[!ht]
\centerline{\psfig{file=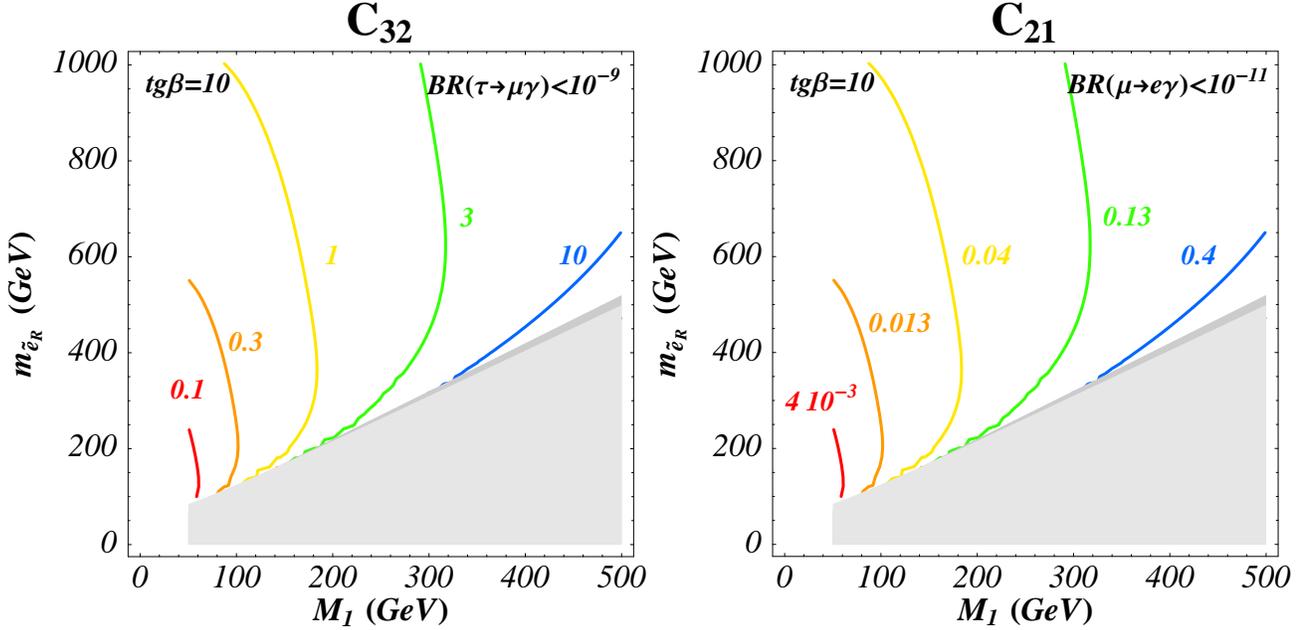,width=1.1\textwidth}}
\caption{Upper limits on $C_{32}$, $C_{21}$ in the plane $(M_1, m_{\tilde e_R})$, respectively the 
$\tilde B$ and right slepton masses. Gaugino and scalar universalities, $\mu$ from radiative electroweak symmetry breaking and 
$m_0=A_0$ have been assumed.}
\label{FR}
\end{figure}

It is well known that, even if $\delta^{(0)}$ vanishes, which is case for universal and real boundary conditions
at $\Lambda$, $\delta^{(rad)}$ can be {\it generated from RGE} if below $\Lambda$ there are new interactions 
beyond those of the SM
\footnote{In GUTs, for instance, $y_t = O(1)$ induces fermion-sfermion misalignement \cite{barhall}: for $SU(5)$ it is however
too small to lead to observable consequences, while for $SO(10)$ it could be sufficient \cite{romstr}.}.
It has been shown \cite{bormas} that the see-saw (even without a stage of grandunification) could lead
to observational consequences, due to a large Yukawa coupling in the Dirac term $y_3 = O(1)$.
Fermion-sfermion misalignement reads
\beq
{\bar m}^2_{\tilde f} \delta^{LL}_{ij}= {m^2}^{LL}_{ij}= \frac{1}{8 \pi^2} (3 m_0^2 + 2 A_0^2) 
\underbrace{(y_\nu^\dagger \ln(\frac{\Lambda}{\hat M}) y_\nu)_{ij}}_{\equiv C_{ij}}~~,
\eeq
where universality has been assumed and $m_0$, $A_0$ are respectively the soft scalar masses and the trilinear
coupling.  
The bound on $\delta^{LL}_{ij}$ of the previous section is then translated into a bound on $C_{ij}$ as displayed 
in fig. \ref{FR}.

It has been shown \cite{simple} that some simple realisations of the see-saw mechanism (for instance some 
models with $U(1)$ flavour symmetry, or with $R=\mathbb{I}$) lead to $\mu \rightarrow e \gamma$, 
$\tau \rightarrow \mu \gamma$ at hand, some models having been already ruled out by 
$\mu \rightarrow e \gamma$. It has been claimed that there exist a certain correlation
between the solar solution and the rate of $\mu \rightarrow e \gamma$, which would be enhanced in the case
of large solar mixing angle.

We addressed the question \cite{lms1, lms2} whether the prediction of high rates is or not 
a general feature of the see-saw.
It turns out that this is not the case and that in general there is no connection between having a large solar angle
and an enhancement of the $\mu \rightarrow e \gamma$ rate. Anyway, it is possible to link the $\tau \rightarrow
\mu \gamma$ rate to physical characteristics of the see-saw: namely it is possible to have some indirect indications
on the possible magnitude of the largeness of the 23 mixing angle of $V_L$, $\theta^L_{23}$,
which, as previously explained, is totally unaccessible from $m_\nu^{eff}$.
That $C$ provide some access to $V_L$ can be easily seen by noticing that, due to the mild effect
of the logarithm, $C$ is in first approximation diagonalised by $V_L$ itself. The more the mixings in $V_L$ are large,
the more the rates will be enhanced.
So, if $y_3 = O(1)$, which we assumed in all the following discussion, 
non observation of $\ell_i \rightarrow \ell_j \gamma$ can 
be translated into an upper bound on the combination ${V_L}_{i3} {V_L}_{3j} \ln(\Lambda/M_3)$.

Despite $m_\nu^{eff}$ provides no direct access to $V_L$, it is possibile to establish an indirect correspondence
by considering the following useful classification criterion for see-saw mechanisms \cite{lms1}.
Consider the hierarchical case for $\nu$ spectrum (for a discussion on the inverted hierarchical see \cite{lms2}).
One can classify every possible see-saw in three cathegories according to which right-handed neutrino
dominates - in the sense explained previously - the scale $m_{atm}$. 
There are obviously three possibilities.

\noindent {\bf None:} all the elements of the third row of $R$ are of comparable magnitude.
This class has the following features: $\theta_{atm} \sim \theta^L_{23}$; the scale of $M_3$, the heaviest $\nu^c$
has to be below $5 \cdot 10^{14}$ GeV; $y_\nu$ is "lopsided"; a tuning of the order of 
$r$ is necessary in order to have at the same time large atmospheric mixing and hierarchical masses. 
All the see-saw models which follow from a $U(1)$ flavour symmetry with charges of
the same sign belong to this class. 
Indeed, it is well known that in this case the actual values of the $\nu^c$ charges
are not important because $m_\nu^{eff}$ depends only on those of $\nu$. This physical information is
well encoded in $R$: its elements are all of the same order but the absence of structure in $R$ has to be
payed by doing the tuning.

\noindent {\bf The heaviest, $M_3$:} $r R_{33} \ge R_{31}, R_{32}$. 
In this case $\theta^L_{23}\approx \theta_{atm}$ with corrections at the level of $r$; again 
$M_3 < 5 \cdot 10^{14}$ GeV and $y_\nu$ is "lopsided", but now we have naturally large atmospheric mixing and 
large mass splittings, so that no tuning has to be introduced. These models necessitate of a 
richer flavour symmetry than those above, for instance a $U(1)_F$ flavour symmetry with positive and negative 
charges, which allow for holomorphic zeros in the textures.

\noindent {\bf One among the lightest, $M_1$ or $M_2$:} $r R_{31(32)} \ge R_{33}, R_{32(31)}$. 
The relevant feature of this class is that $\theta^L_{23}$ is no more linked to $ \theta_{atm}$.
On the contrary, $y_\nu$ can possess small mixings so that $\theta^L_{23}$ could even vanish.
For this class $M_3 \ge r^{-1} 5 \cdot 10^{14}$ GeV - which is quite good for $SO(10)$ -
large atmospheric mixing and 
large splitting is naturally realised. These interesting models necessitate of an even
richer flavour symmetry than those above: models have been studied with several $U(1)_F$'s 
with positive and negative charges and with non-abelian flavour symmetries.

The remarkable fact that emerges from this classification is that the first two classes 
require $\theta^L_{23} \sim \pi/4$, so that $C_{32} \sim \ln(\Lambda/ M_3) \sim 7$, while for 
the third class $C_{32}$ can be much smaller.
It is thus sufficient to give a look to fig. \ref{FR} to understand that future searches 
for $\tau \rightarrow \mu \gamma$ have the potentiality of discriminating between these cathegories. 
The present limit on $BR(\tau \rightarrow \mu \gamma)$ implies a limit on $C_{32}$ weaker by a factor of $30$
with respect to fig. \ref{FR}, so that no conclusion can be drawn by now.

For $\mu \rightarrow e \gamma$ the situation is not so sharp:
for the last two cathegories of models, dominance of the heaviest and of one among the lightest, 
the predictions are very model dependent. Instead, the models with only positive $U(1)_F$ charges
of the first class predict $C_{21} \sim 7~r$: if supersymmetry is somewhere at left of the green curve, 
such models are already excluded for $r$ in the range of LMA. 
It is worth to stress that, if $\mu \rightarrow e \gamma$ were not found, this would be perfectly 
compatible with the two last cathegories, maybe slighly favouring the third one.
Turning the argument around we can notice that  $\mu \rightarrow e \gamma$ 
could test the product $V_{13}^L V_{23}^L$ up to the level of 
$\sim .005$, which corresponds the measured value of the analogous mixings in $V_{CKM}$. 
Notice also that anarchical models, for which $C_{21} \sim 7~y^2$ ($y$ being the common Yukawa),
are already in crisis in all the supersymmetric plane.

It is also interesting to ask if the planned experimental improvement in the sensitivity to $d_\mu$ will 
imply significant progresses also from the model building point of view.
This is the case because, even starting with universal and real boundary condition at $\Lambda$, 
$\delta^{(rad)}$ could induce an EDM. In the case of the see-saw, however, 
if $\hat M \propto \mathbb{I}$, the naive scaling $d_e/d_\mu \sim m_e/m_\mu$ - 
which, given the present bound on $d_e$, would imply $d_\mu < 10^{-25}$ e cm - cannot be avoided.
On the contrary, in case of hierarchical $\nu^c$, the naive scaling can be avoided
due to the FC contribution \cite{ellisetal}.
In any case, the FV part is such that the phases in the four products of 
$\delta$'s in eq. (\ref{4im}) are very small \cite{ms_edm}: this means that if the SM is extended 
to just the see-saw, then $d_\mu^{FV}$ is always smaller than the maximum value allowed from 
non observation of  $\tau \rightarrow \mu \gamma$.

\vskip 0.5 cm
\noindent {\bf Acknowledgements} I would like to thank Carlos Savoy for pleasant discussions 
and collaboration.




\end{document}